\documentclass[twocolumn,showpacs,preprintnumbers,amsmath,amssymb]{revtex4}
                                                                                
\usepackage{graphicx}
\usepackage{dcolumn}
\usepackage{subfigure}
\usepackage{bm}
\def\be{\begin{equation}}
\def\ee{\end{equation}}
\def\bc{\begin{center}}
\def\ec{\end{center}}
                                                                                
\begin{document}
                                                                                
\preprint{CUPhys/03/2007}
                                                                                
\title{Multidimensional persistence behaviour in an Ising system}
                                                                                
\author{Anjan Kumar Chandra$^1$}
\author{Subinay Dasgupta}%
\affiliation{%
Department of Physics, University of Calcutta,92 Acharya Prafulla Chandra
Road, Calcutta 700009, India.\\
}%

\date{\today}
                                                                                
\begin{abstract}
We consider a periodic Ising chain with nearest-neighbour and $r$-th neighbour
interaction and quench it from infinite temperature to zero temperature.
The persistence probability $P(t)$, measured as the probability that a spin
remains unflipped upto time $t$, is studied by computer simulation for
suitable values of $r$. We observe that as time progresses,
$P(t)$ first decays as $t^{-0.22}$ (-the {\em first} regime), then the $P(t)-t$
curve has a small slope (in log-log scale) for some time (-the {\em second}
regime) and at last it decays nearly as $t^{-3/8}$ (-the {\em third} regime). 
We argue that in the first regime, the persistence behaviour is the usual one
for a two-dimensional system, in the second regime it is like that
of a non-interacting (`zero-dimensional') system and in the
third regime the persistence behaviour is like that of a one dimensional
Ising model. We also provide explanations for such behaviour.
\end{abstract}

                                                                                
\def\be{\begin{equation}}
\def\ee{\end{equation}}
\maketitle

\section{Introduction} 

\footnotetext[1]{{\em Present address} : Theoretical Condensed Matter Physics Division and Center for Applied Mathematics and Computational Science, {\em Saha Institute of Nuclear Physics}, 1/AF Bidhannagar, Kolkata 700064, India}
   The tendency of a spin in a spin-$\frac{1}{2}$ Ising system to remain in 
its original state following a quench from infinite temperature to zero 
temperature has been extensively studied over the last decade and is an 
example of the phenomenon called persistence in dynamical systems 
\cite{Majumdar}-\cite{derrida2}. The probability $P(t)$ that a spin does not 
flip upto time $t$, exhibits a power law behaviour 
\be P(t) \sim t^{-\theta},  \label{eq:5.1}\ee 
where $\theta$ is a non-trivial exponent, as it is not related to any other 
static or dynamic exponent. In one dimension with nearest-neighbour 
interaction, it has been proved exactly that this exponent is $\theta=3/8$ 
\cite{derrida1}. In two dimensions, again with nearest-neighbour interaction, 
this exponent has been numerically evaluated \cite{jain} as 
$\theta=0.209$. For Glauber dynamics, the one-dimensional Ising system is 
equivalent to a one-dimensional $A + A \rightarrow 0$ diffusion system 
(see below). Hence, the ``zero-dimensional'' persistent behaviour should 
correspond to that of a system of 
non-interacting particles initially spread randomly over a chain with 
density $\rho$
and then diffusing independent of each other. (The rule of diffusion is to
take a step to the right
or to the left with probability $1/2$.) The persistence probability for
this system has been shown \cite{bray} to decay stretched-exponentially, 
\be P(t) = \exp[-(2\sqrt{2/\pi}\rho)\,\sqrt{t} ] \label{eq:5.2} \ee
The objective of this communication is to report the observation that an Ising
system with nearest-neighbour interaction on a rectangular helical lattice,
when quenched to zero-temperature from an infinite temperature shows 
zero-, one- and two-dimensional persistence behaviour in different regions of
its temporal evolution. In the next section, we shall describe the details of 
the system simulated and the algorithm followed, alongwith the results. 
An explanation of the
simulation results will also be presented in Sec. III. In the last section 
we shall discuss some subtle issues.

\begin{figure}[t]
\bc
\noindent \includegraphics[clip,width= 6cm, angle = 0]{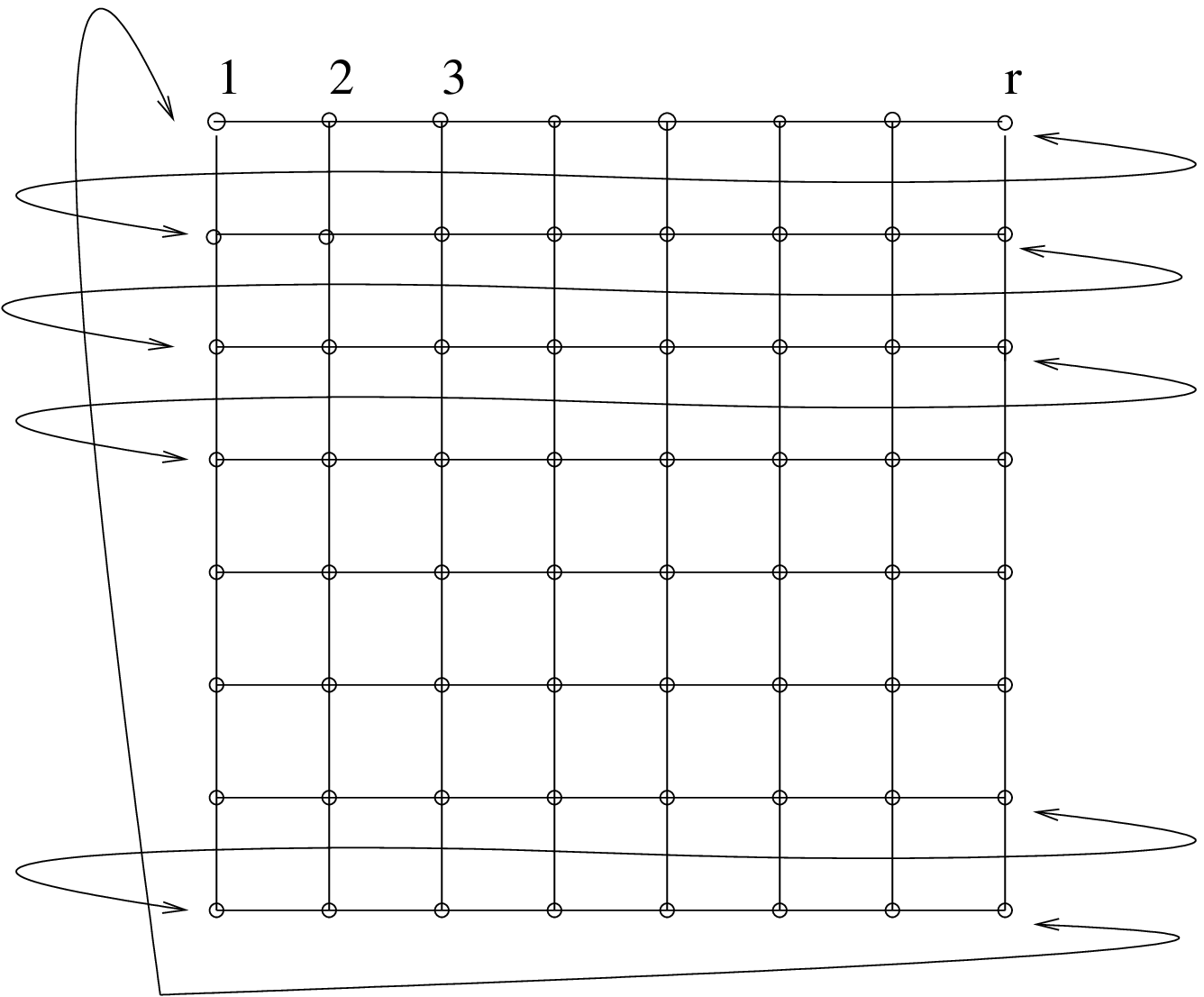}
\caption\protect{\label{fig:helical}
The lattice with helical boundary condition.}
\ec
\end{figure}

\section{The algorithm and the results } 
In this section we shall first describe the details of our system and our
algorithm, and then present the results.
Consider Ising spins $s_1$, $s_2$, $\cdots$, $s_L$ placed on a chain 
of $L$ sites with the periodic boundary condition $i+L \equiv i$. Initially, 
each spin is $+1$ or $-1$ with equal probability. One iteration of the system 
consists of the following 4 steps: \\
(i) choose one (say, $k$-th) spin randomly, \\
(ii) calculate its energy 
\be E_k \equiv s_k(s_{k-1} + s_{k+1} + s_{k-r} + s_{k+r}), \label{eq:5.3}\ee \\
(iii) flip $s_k$ with probability 1 if $E_k > 0$, and with probability 
$\frac{1}{2}$ if $E_k = 0$ (do not flip at all if $E_k < 0$), \\
(iv) repeat the steps (i) to (iii) $(L-1)$ times more (random updating). \\
Here, $r$ is a parameter of the model and must lie between $2$ and $L$. Clearly,
we have nearest and $r$-th neighbour interaction and
our system is equivalent to a nearest neighbour rectangular Ising system 
of size $r \times (L/r)$. The lattice is {\em not} periodic in the two
axial directions, rather the chain is wound as a helix of periodicity $r$ 
with the ends (first and $L$-th sites) put side by side 
(Fig.~\ref{fig:helical}). When 
$r \ll \sqrt{L}$, the system is effectively a one-dimensional one, while
for $r \sim \sqrt{L}$ it is a two-dimensional one with aspect ratio 
$a=r^2/L$. In this communication we study the case of $a \sim 0.01$, so that 
the system is effectively one-dimensional, or at least a narrow strip. 
We do not consider the situation when the condition $r \ll \sqrt{L}$ is not
satisfied. 

\begin{figure}
\bc
\noindent \includegraphics[clip,width= 6cm, angle = 270]{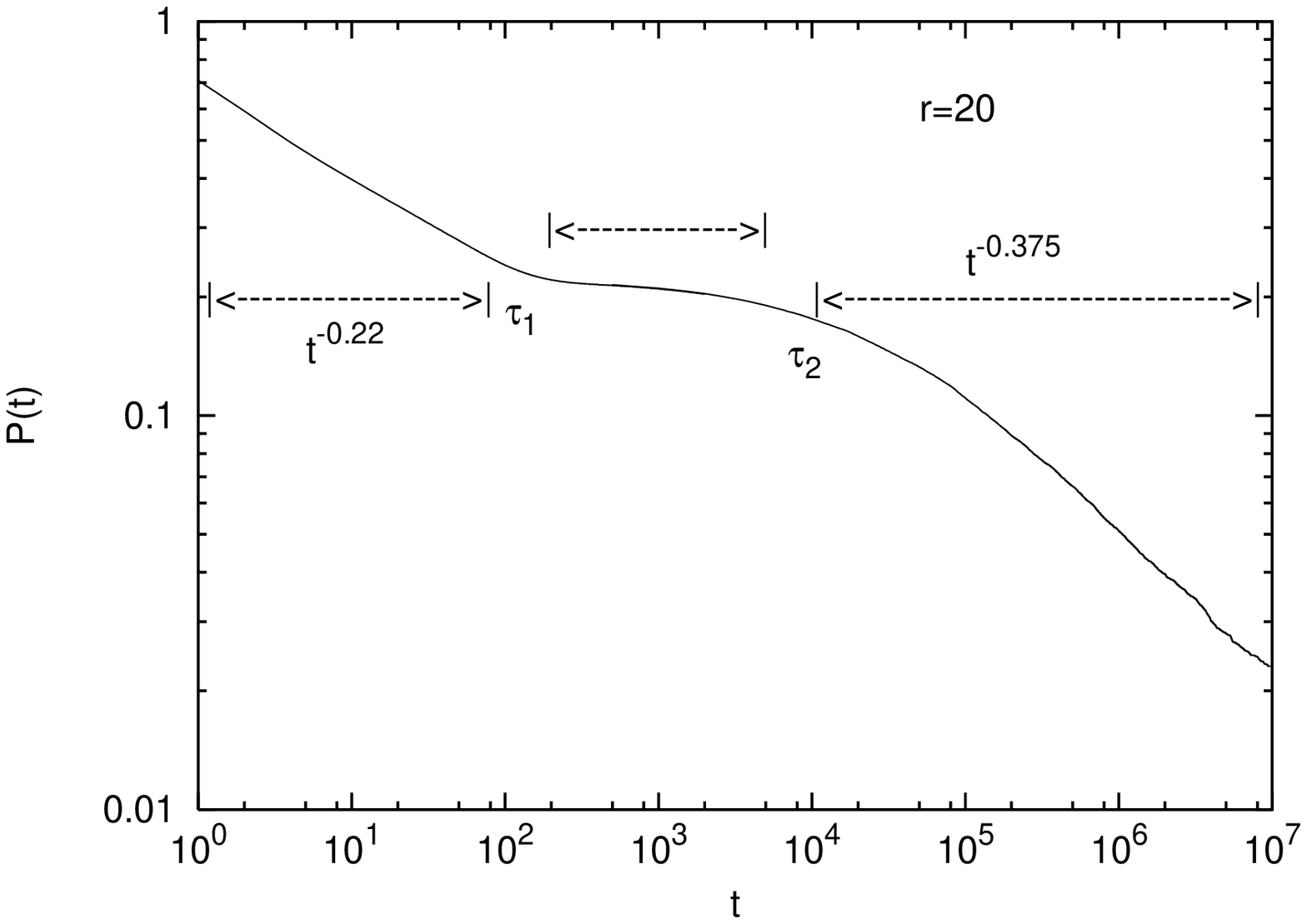}
\caption\protect{\label{fig:Phaseran}Plot of P(t) as a function of
$t$ for $r=20$. The curve is for
$L=10000$ and averaged over $100$ configurations. The first, second and third
regime are marked in the figure. The curves for $L=5000$ and $L=20000$ fall on
the curve shown here.}
\ec
\end{figure}

\begin{figure}
\bc
\noindent \includegraphics[clip,width= 6cm, angle = 270]{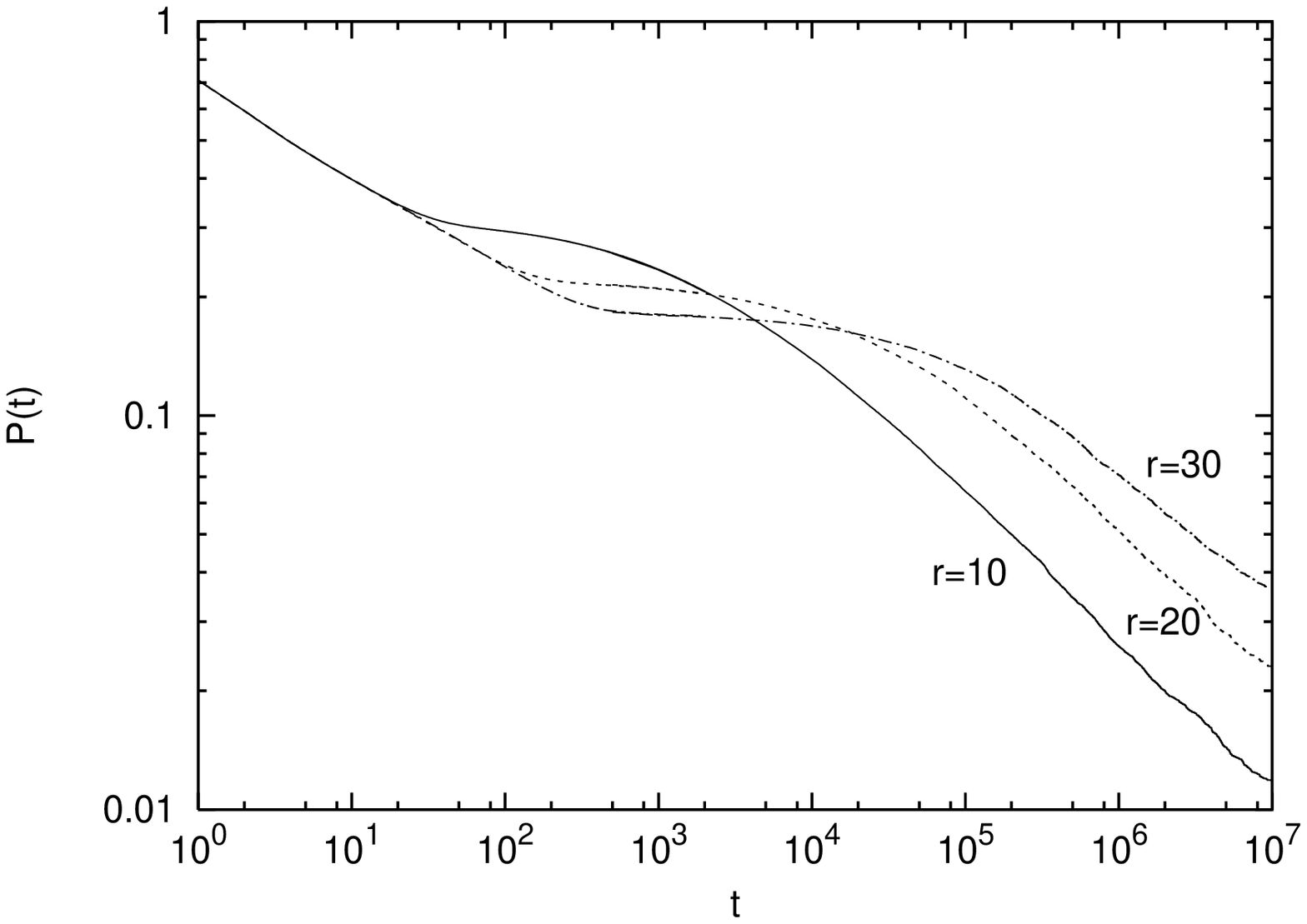}
\caption\protect{\label{fig:rr2ran.L10e3}Plot of P(t) as a function of
$t$ for $r=10, 20$ and $30$. The curve is for
$L=10000$ and averaged over $100$ configurations.}
\ec
\end{figure}

\begin{figure}
\bc
\noindent \includegraphics[clip,width= 6cm, angle = 270]{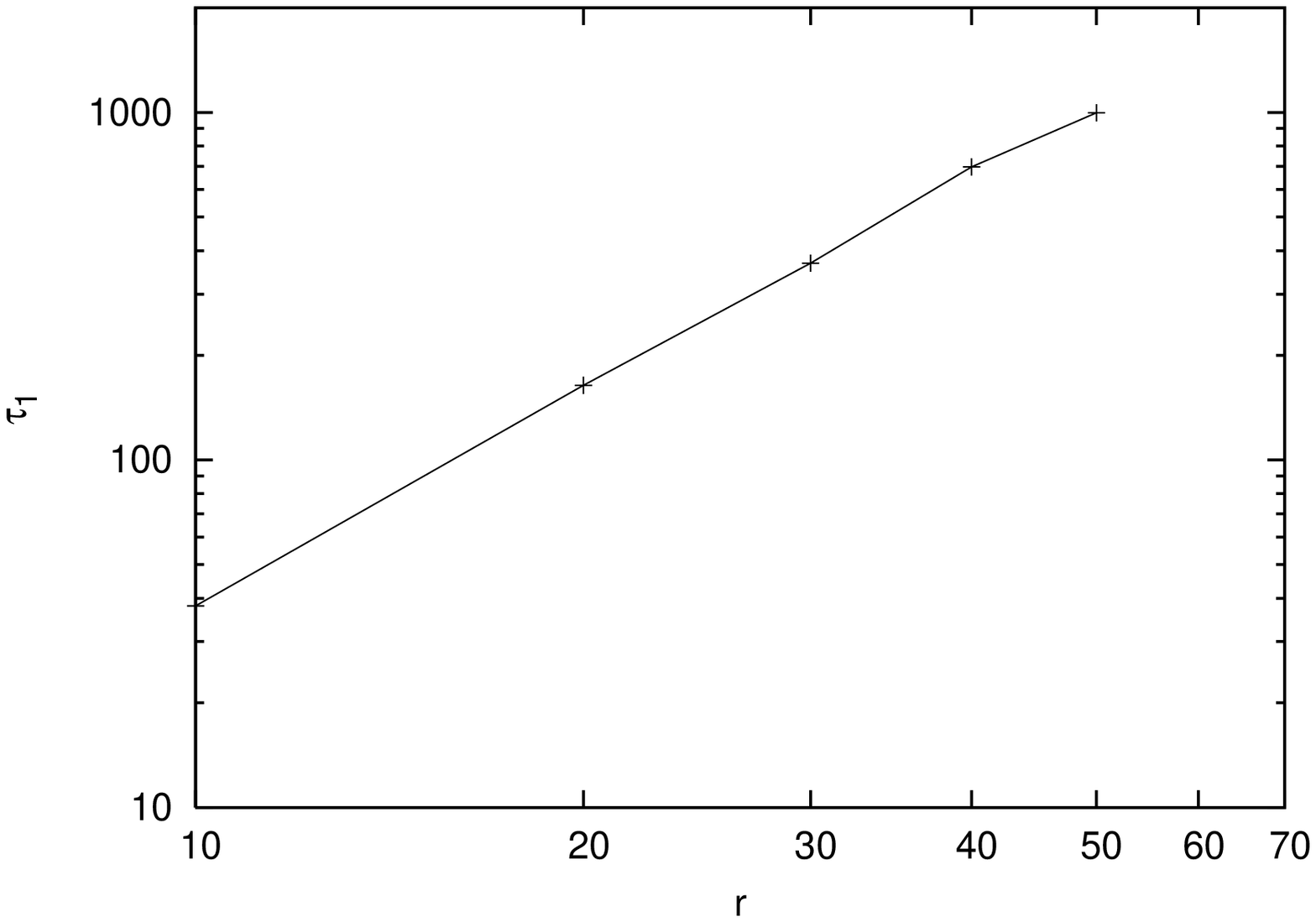}
\caption\protect{\label{fig:Tpvar}Variation of $\tau_1$ vs $r$. The curve
is for $L=10000$ and averaged over $100$ configurations.}
\ec
\end{figure}
                                                                                
\begin{figure}
\bc
\noindent \includegraphics[clip,width= 12cm,angle = 0]{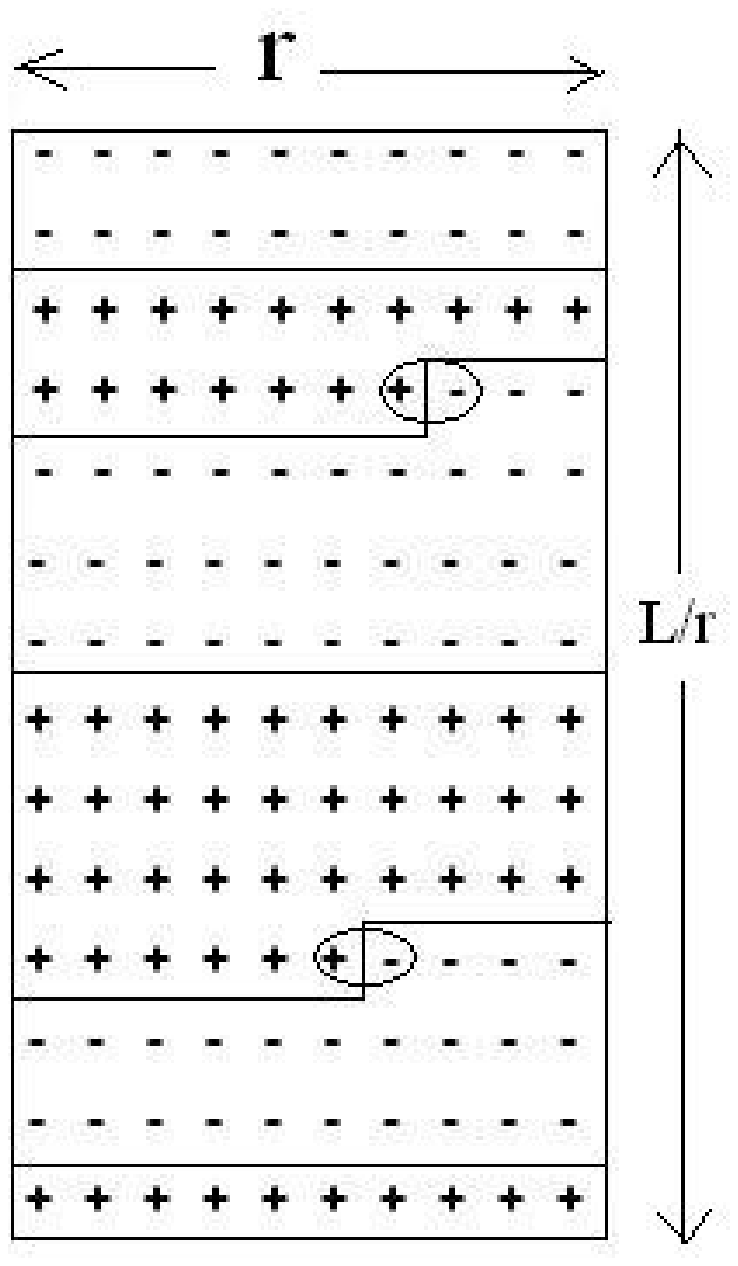}
\caption\protect{\label{fig:Latt2}Presence of only a small number of kinks
in the domain wall ($A$ particles) in the second regime. Dynamics occurs only
at the kinks.}
\ec
\end{figure}

After every iteration we compute the fraction of spins that has not been 
flipped till now. This fraction, averaged over many realisations of the system
gives us the probability of persistence $P(t)$.
We present in Figs.~\ref{fig:Phaseran} and ~\ref{fig:rr2ran.L10e3} the 
simulation results for this quantity. 
At first $P(t)$ decreases as $t^{-\theta}$ with $\theta=0.22 \pm 0.01$ showing 
two-dimensional behaviour. This behaviour continues upto a certain
time, say, $\tau_1$. For a given $L$, the value of $\tau_1$ has been observed
to increase with $r$ as $r^2$ (Fig.~\ref{fig:Tpvar}). 
We call the region $1 < t < \tau_1$ the {\em first regime}. Next follows 
the {\em second regime}
extending upto some $\tau_2$ iteration where the $P(t) - t$ curve has a small
slope in the log-log scale. 
We shall see below that in this regime the system behaves as a 
zero-dimensional one. For limitation of computational resources, we could
not achieve precise evaluation of $\tau_2$, but could observe that, like
$\tau_1$, this quantity increases with $r$.
At last comes the {\em third regime} (for $\tau_2 < t < \infty$) where $P(t)$ 
decays nearly as $t^{-\theta}$ with $\theta=0.375 \pm 0.01$ 
showing the one-dimensional behaviour. 

\section{Explanation of the results}

We now explain the observations in the 
three regimes one by one, by approximate analytic arguments but 
an exact analytic calculation of the persistence behaviour spanning 
over the three regimes is yet to be done.

\begin{figure}
\bc
\noindent \includegraphics[clip,width= 6cm, angle = 270]{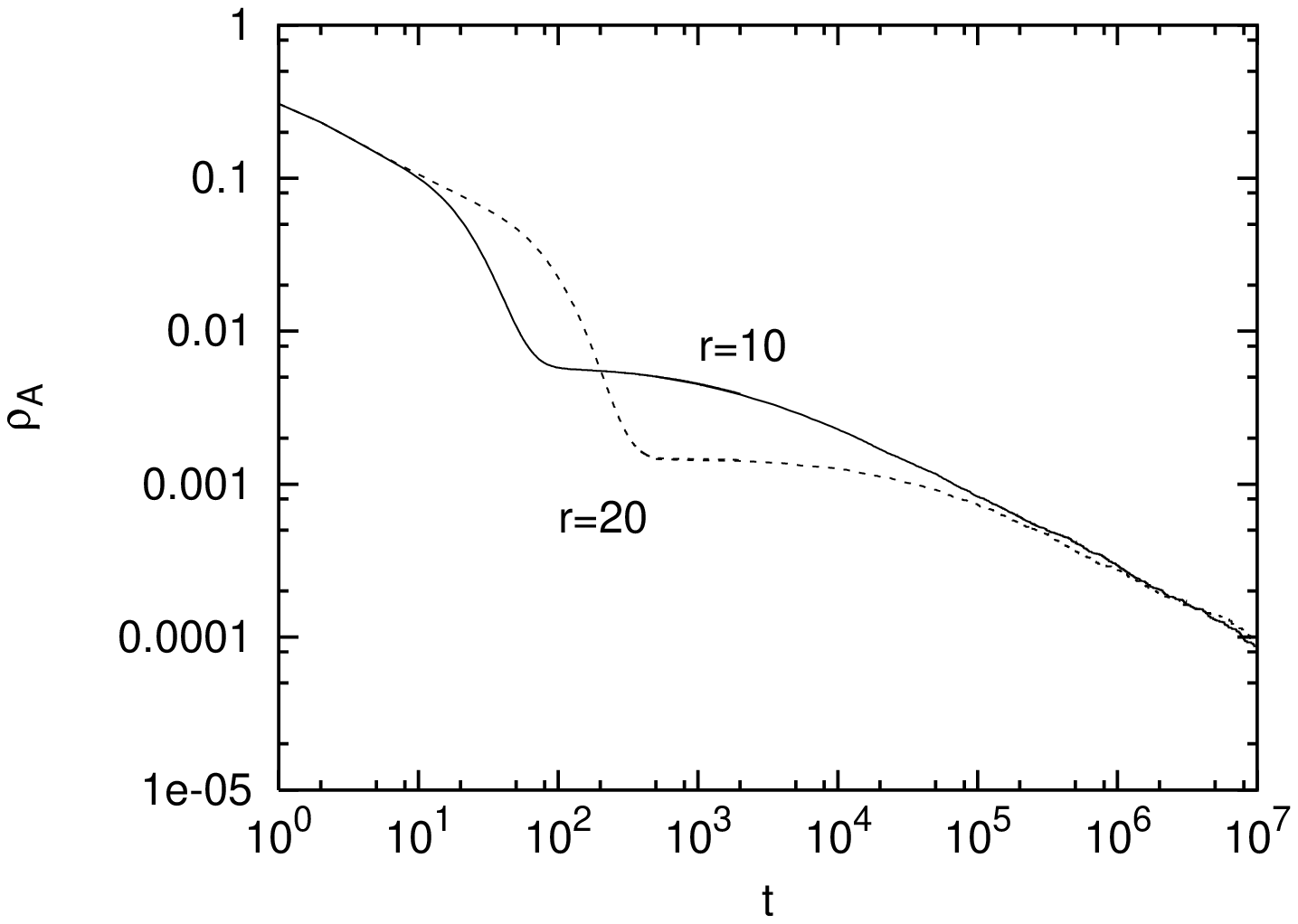}
\caption\protect{\label{fig:ntran.k10}Plot of density $\rho_A$ of $A$ 
particles as a function of $t$ for $r=10$ and $20$, $L=10000$. The data was
averaged over $1000$ configurations. In the nearly horizontal region (the second
regime) $\rho_A=5.50 \times 10^{-3}$ for $r=10$ and 
$\rho_A=1.44 \times 10^{-3}$ for $r=20$.}
\ec
\end{figure}

\begin{figure}
\bc
\noindent \includegraphics[clip,width= 6cm, angle = 0]{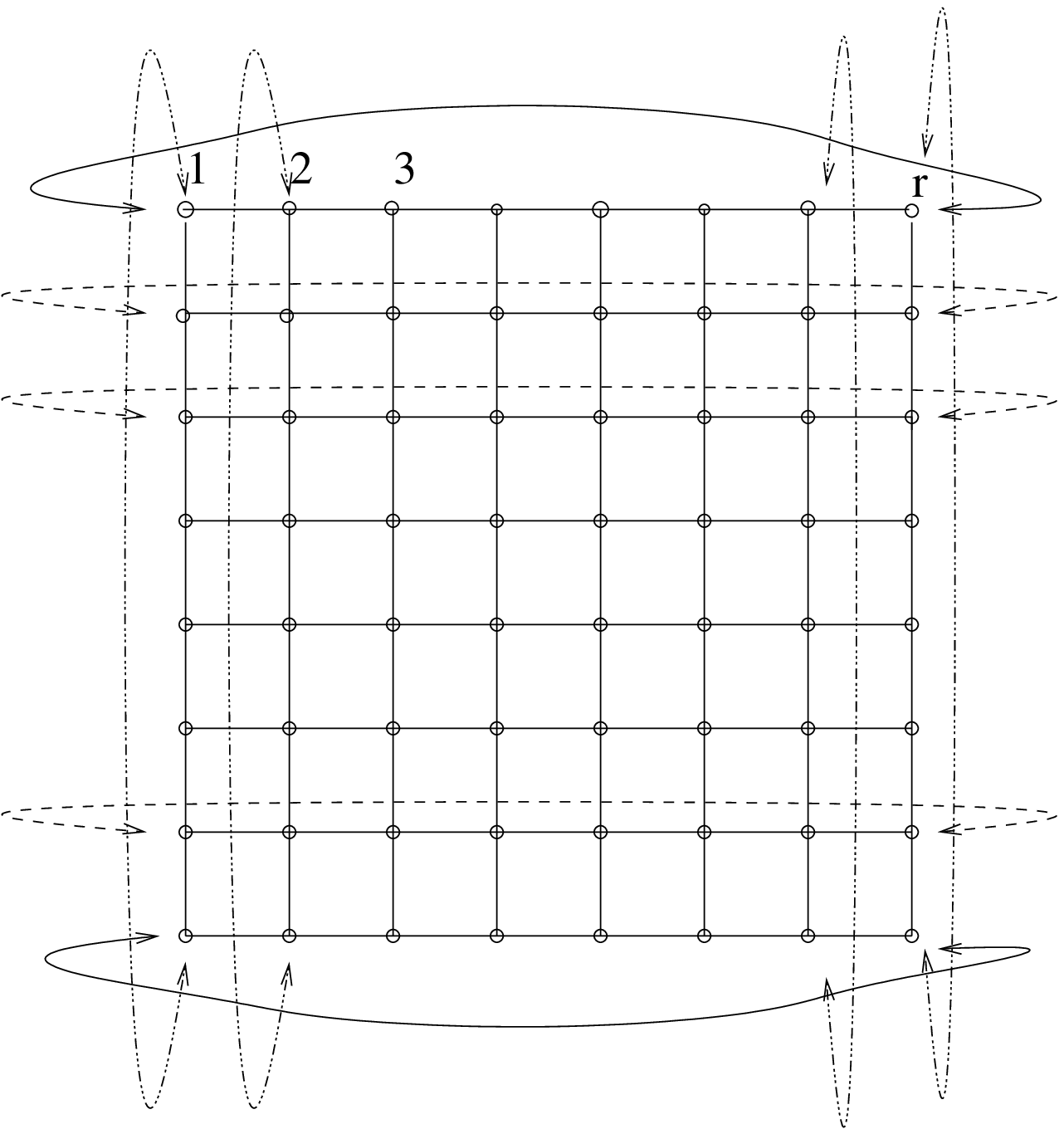}
\caption\protect{\label{fig:periodic}The lattice with periodic boundary
condition.}
\ec
\end{figure}

{\em The First Regime} : Here 
the system shows normal two-dimensional behaviour. The
persistence curve $P(t)$ vs. $t$ saturates at a time $\tau_1 \sim r^z$ to a 
value that varies as $r^{-z\theta}$ where $z$ is the dynamical critical 
exponent ($\approx 2$) \cite{Majumdar, Ray}. 
After the system reaches the saturation stage, the domain walls (lines 
separating unlike spins) are mostly parallel to the X axis, with only a small 
number of kinks (Fig.~\ref{fig:Latt2}). The dynamics occurs only at these 
kinks, which we call `$A$ particles'. For the updating rule stated above, 
each $A$ particle jumps to the left or right with probabilities
1/2, 1/2, assuming that the walls are sparsely distributed
over the system (an $A$ particle does not have another at a distance $\le r$). 
The density $\rho_A$ of $A$ particles, measured as the number of vertical 
domain walls per site is shown in Fig.~\ref{fig:ntran.k10}.
This quantity also shows a plateau region in the second regime, like 
persistence.  It is important to note that, for a fully periodic system shown 
in Fig.~\ref{fig:periodic}, there can be only an even number of $A$ particles 
in a row. Presence of one particle in a row is hence ruled out and 
two or more particles get annihilated within time $\tau_1$. The dynamics
therefore {\em stops completely} at $t = \tau_1$ in the case of a fully 
periodic system, and the second and the third regimes do not appear.

\begin{figure}[h]
\bc
\noindent \includegraphics[clip,width= 6cm, angle = 270]{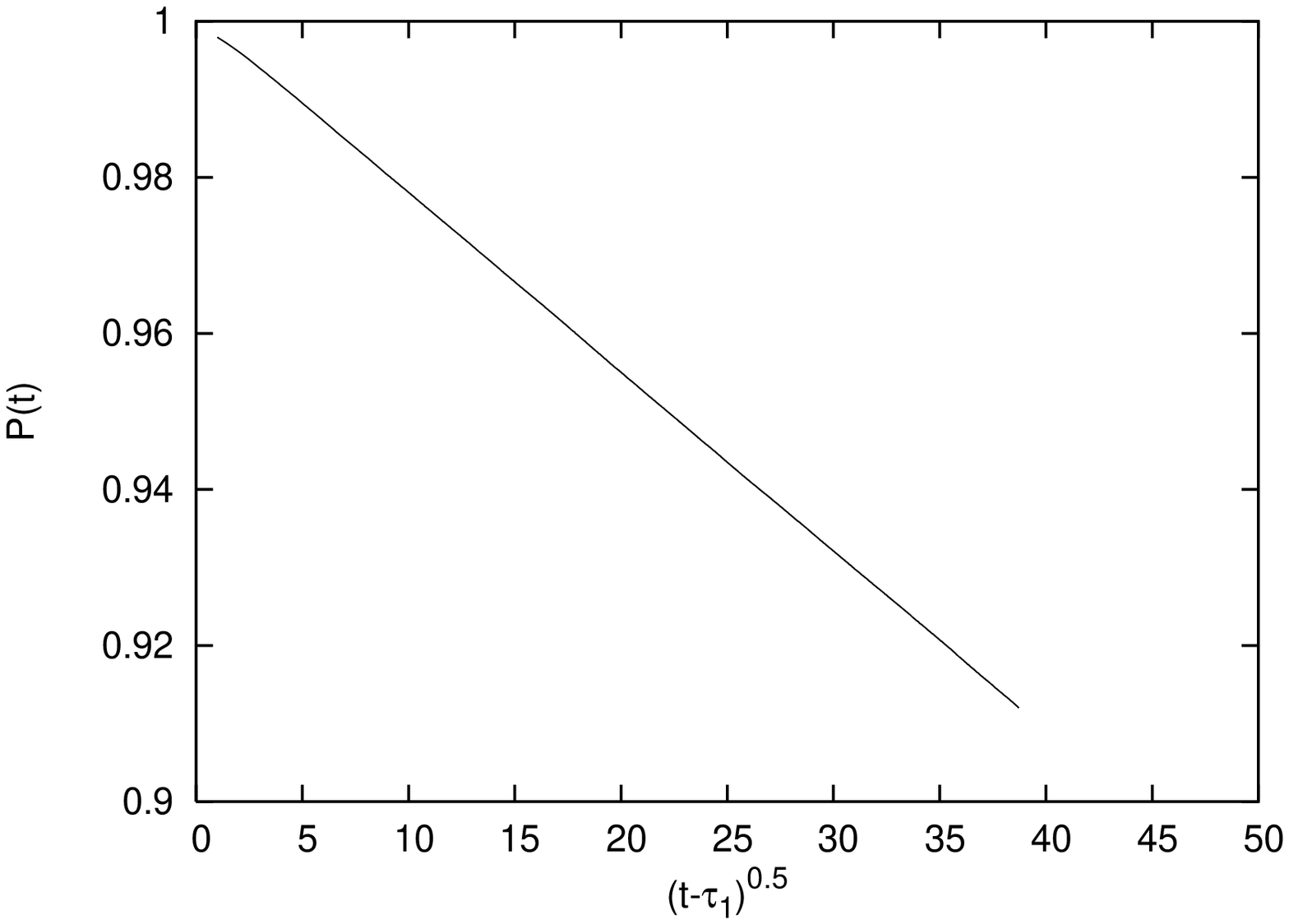}
\caption\protect{\label{fig:secondreg3}Plot of $P(t)$ as a function of
$\sqrt{t-\tau_1}$ in the second regime of the persistence curve for
$r=20$ and $L=10000$. $P(t)$ was set to be 1 at $\tau_1=500$. The curve
fits to $1-1.58\rho_A\sqrt{t-\tau_1}$ with $\rho_A=1.44 \times 10^{-3}$.
The data were averaged over $1000$ configurations.}
\ec
\end{figure}

{\em The Second Regime} : 
What happens to our (helical) system for $t > \tau_1$ ? For an interval of time 
$\tau_1 < t < \tau_2$, the $A$ particles (kinks in the domain walls) are 
so far away from each other that they cannot ``see'' each other and diffuse 
independently. For a system of independent random walkers of density $\rho$, 
the persistence has been shown \cite{bray} to obey Eq.~\ref{eq:5.2}. To 
compare our
persistence data with Eq.~\ref{eq:5.2}, we assign each site to be persistent at 
$t = \tau_1$. This makes $P(\tau_1)=1$ and obliterates the distribution of 
persistent sites created in the first regime. Then we note down the (almost 
constant) density $\rho_A$ in the second regime and observe that the $P(t)$
data here obeys the relation 
\be P(t) = 1 - \alpha\rho_A\sqrt{t-\tau_1} \label{eq:5.4} \ee 
with $\alpha \approx 1.6$ (Fig.~\ref{fig:secondreg3}). 
Since the value of the slope $\alpha$ is close to $2\sqrt{2/\pi} = 1.596$, and
since the value of $\rho_A$ is small, 
Eq.~\ref{eq:5.2} is obeyed and we conclude that 
in the second regime, the system behaves as one of zero-dimension.

\begin{figure}
\bc
\noindent \includegraphics[clip,width= 6cm, angle = 270]{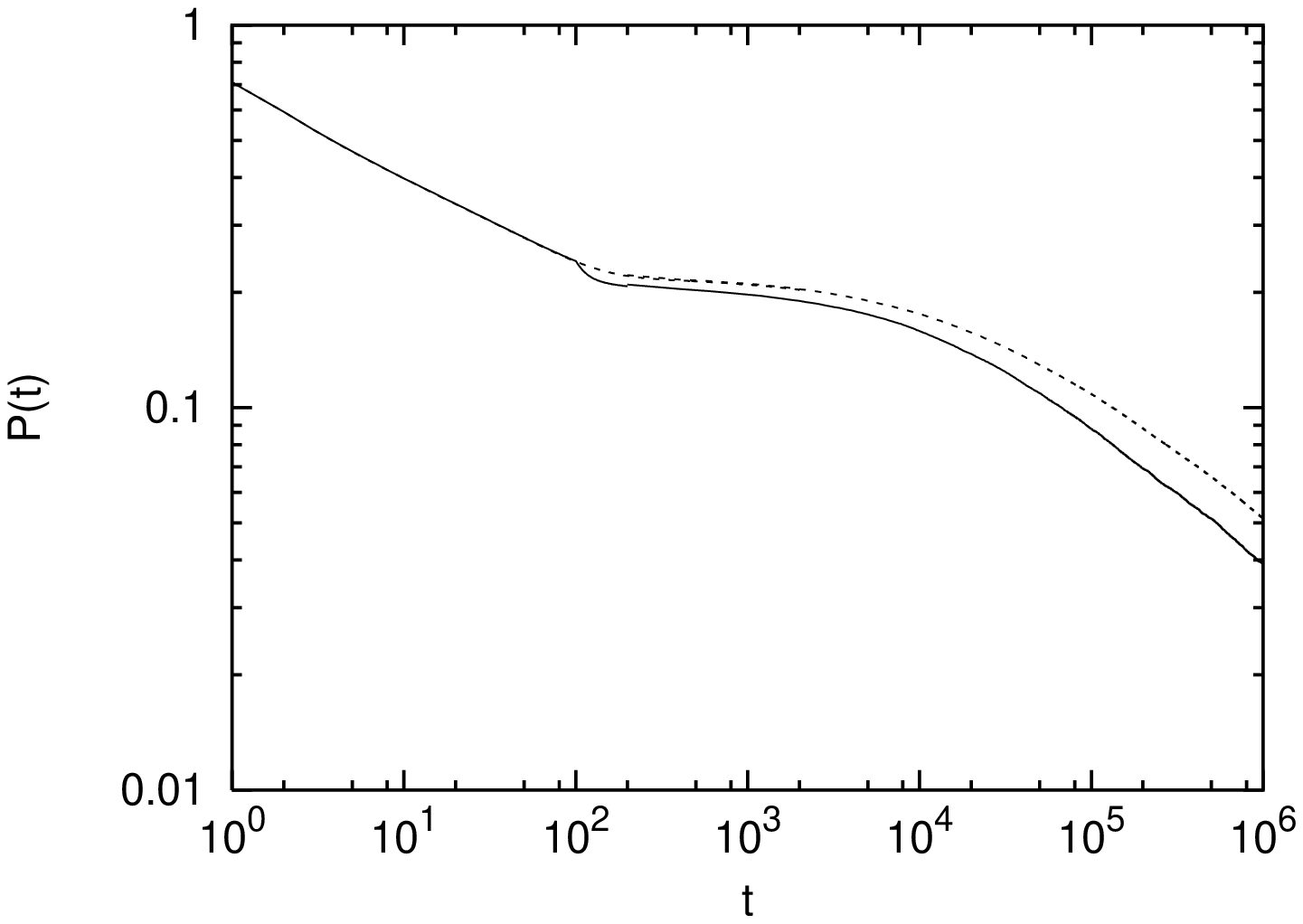}
\caption\protect{\label{fig:ranlongoff}Plot of P(t) as a function of
$t$ for simultaneous nearest neighbor interaction and long-range interaction
$r=20$. After 100 iterations the long-range interaction
is switched off (continuous line). The curve is for $L=10000$ and averaged 
over $1000$
configurations. The dotted line corresponds to the usual case, when the $r$-th
neighbour interaction continues for the entire range of time.}
\ec
\end{figure}

\begin{figure}[h]
\bc
\noindent \includegraphics[clip,width= 5cm, angle = 270]{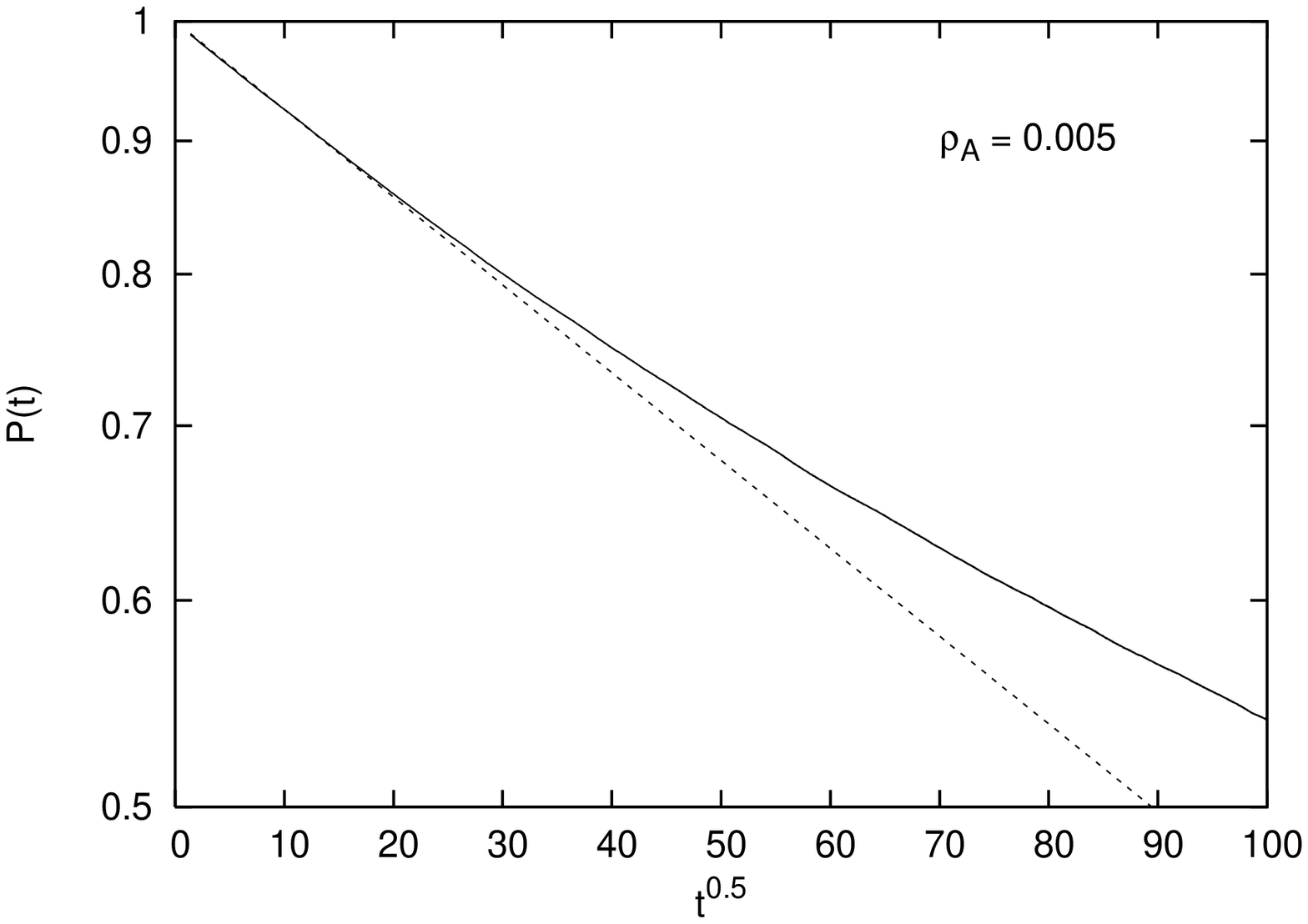}
\caption\protect{\label{fig:aaregime2}Plot of P(t) as a function of $\sqrt{t}$ 
for $A+A \rightarrow 0$ dynamics in log-linear scale. At $t=0$, the $A$ 
particles were randomly spaced with density $\rho_A=0.005$. Here 
$L=4000$ and the results were averaged over $200$ realisations. The initial 
portion ($t<400$) fits to Eq.~\ref{eq:5.2} with $\alpha=1.55$.}
\ec
\end{figure}

\begin{figure}[h]
\bc
\noindent \includegraphics[clip,width= 5cm, angle = 270]{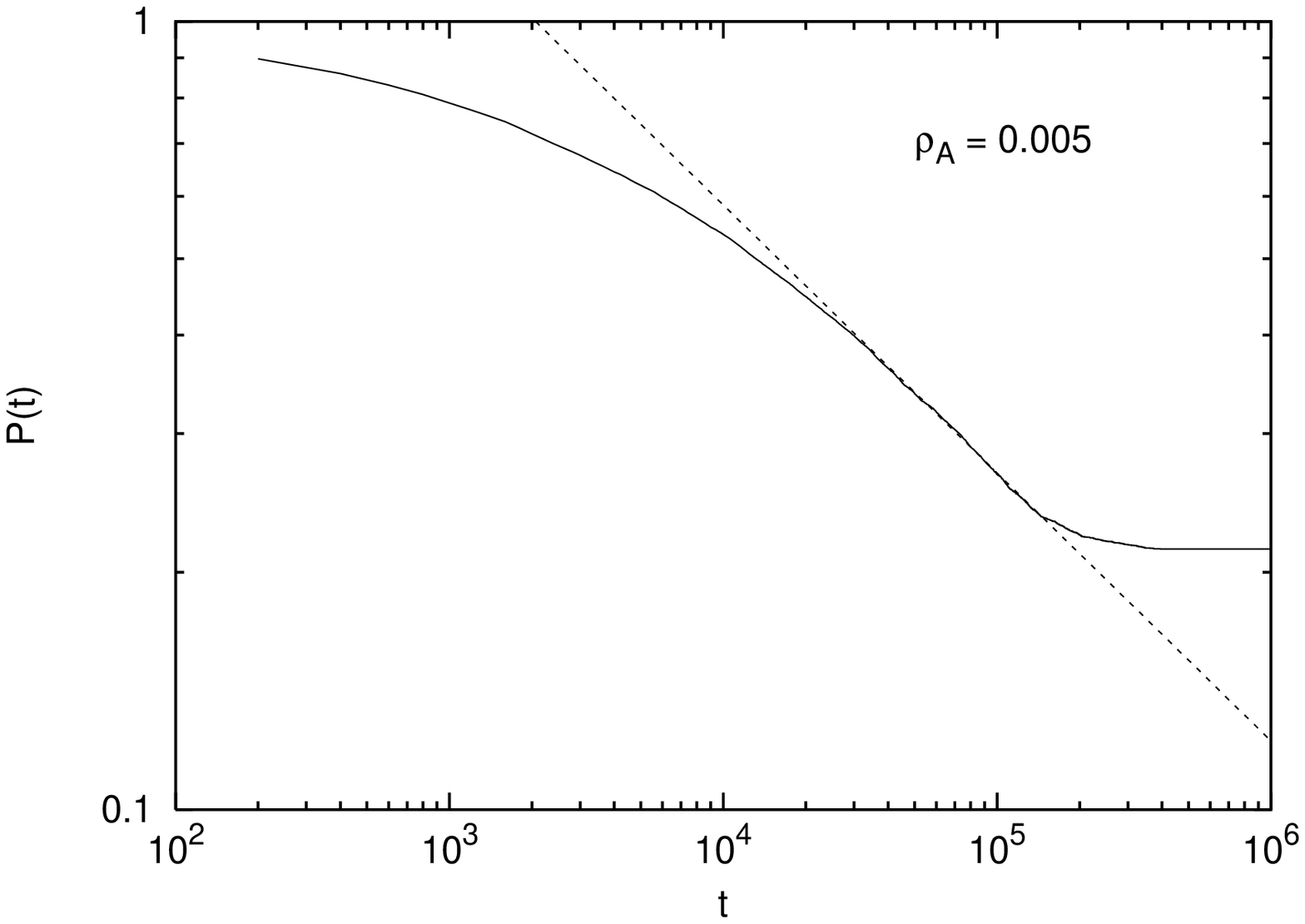}
\caption\protect{\label{fig:aaregime3}Plot of P(t) as a function of $t$ for
$A+A \rightarrow 0$ dynamics in log-log scale. At $t=0$, the $A$ 
particles were
randomly spaced with density $\rho_A=0.005$. Here $L=4000$ and the 
results were averaged over $100$ realisations. The algebraic portion fits
to $13.4t^{-0.34}$}
\ec
\end{figure}

{\em The Third Regime} : 
When the system evolves further ($t > \tau_2$), the $A$ particles start meeting
(and annihilating) each other and the usual one-dimensional dynamics leads to 
$t^{-\theta}$ behaviour with $\theta \approx 3/8$. Since the density $\rho_A$ 
is too small, one needs
to average over a large number of realisations and simulation of the third
regime is a computationally intensive job. The $P(t) - t$ curves remain 
the same for different values of $L$ for a given $r$, but get shifted 
(maintaining $t^{-3/8}$ behaviour) as one varies $r$ at a given $L$ 
(Figs.~\ref{fig:Phaseran} and ~\ref{fig:rr2ran.L10e3}).  
That the dynamics in the second
and the third regimes is indeed described by simple one-dimensional 
$A+A \rightarrow 0$ dynamics is further corroborated by two numerical 
experiments : \\
(i) If we turn off the $r$-th neighbour interaction at $t=\tau_1$, the 
slope of the persistence curve (in log-log scale) does not change much 
(Fig.~\ref{fig:ranlongoff}), 
indicating that it is chiefly the nearest-neighbour interaction that drives 
the dynamics. \\
(ii) Let us consider a periodic chain of $L$ sites and sprinkle randomly
some particles (excluding multiple occupancy at a site) with density $\rho$.
Starting with a low ($\sim 0.005$) value of $\rho$, we let the system 
evolve according to the usual $A+A \rightarrow 0$ dynamics. The result is the 
(zero-dimensional) 
second regime (Fig.~\ref{fig:aaregime2}) for the first 100 iterations,
followed by the (one-dimensional) third regime (Fig.~\ref{fig:aaregime3}). 
The second regime is found to follow Eq.~\ref{eq:5.2} with $\alpha=1.55$ and 
the third regime shows the usual $t^{-3/8}$ behaviour.

\section{Discussion}

(1) We shall first discuss a 
subtle issue regarding the dynamics in the third regime of the Ising model.
When two ``kinks'' of Fig.~\ref{fig:Latt2} come closer than $r$ to each other, 
a domain of length less than $r$ is formed. The rules
of update renders (i) every spin lying within this domain liable to flip (with
probability 1/2) and (ii) the two spins at the ends of this domain bound to
flip (with probability 1). The dynamics thus differs from the one for 
the $A + A \rightarrow 0$ model, since once the domain is
less than $r$ in length, it is swapped within the next $r$ steps. But since we 
are primarily interested in the region $t \gg r$, this difference is not of 
much consequence. One can flip the spins {\em only} at the ends of the domain 
(keeping the ones within the domain unflipped) if we replace the energy
expression of Eq.~\ref{eq:5.3} by
\be E_k \equiv s_k[s_{k-1} + s_{k+1} + \kappa (s_{k-r} + s_{k+r})] \label{eq:5.5} \ee
and choose $\kappa < 1$. We have checked that the persistence behaviour still
remains almost the same.

\begin{figure}
\bc
\noindent \includegraphics[clip,width= 5cm, angle = 270]{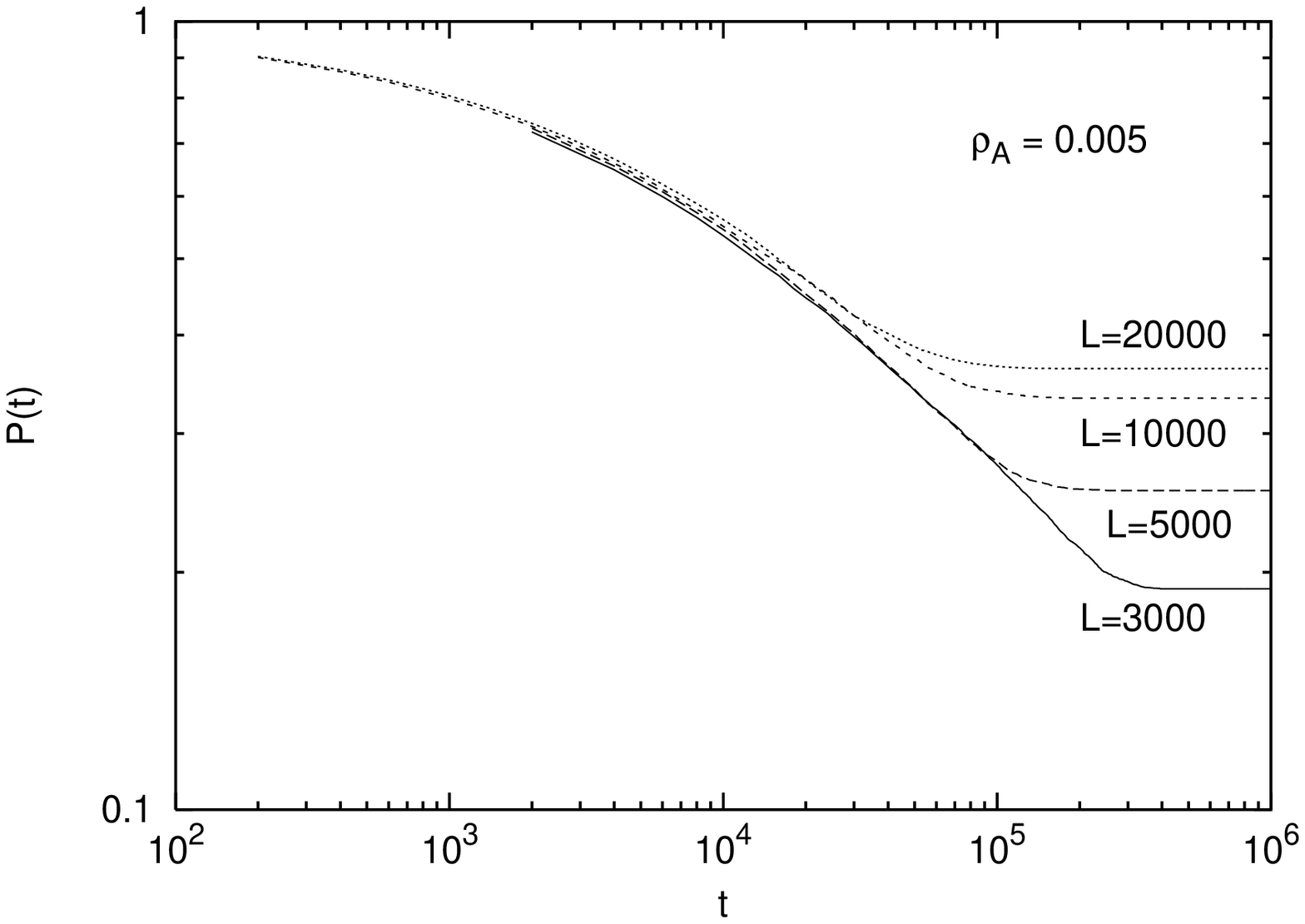}
\caption\protect{\label{fig:a1all}Plot of P(t) as a function of $t$ for
$A+A \rightarrow 0$ dynamics in log-log scale. At $t=0$, the $A$ particles
were randomly spaced with density $\rho_A=0.005$. The size of the samples are
indicated in the figure and the results were averaged over $50$ to $200$ 
realisations. The saturation region rises with increasing system size. Just 
before saturation, the curve does not show algebraic decay over any 
appreciably long region.  }
\ec
\end{figure}

(2) In the case of $A + A \rightarrow 0$ model with initial density
$\rho_A = 0.005$, we
have shown a $t^{-3/8}$ behaviour in Fig.~\ref{fig:aaregime3}. However, one
comes across anomalous behaviour at somewhat larger system size
(Fig.~\ref{fig:a1all}). Thus, after the
third regime continues for some time, one comes across another saturation 
region,
with the saturation value {\em increasing} with increase of $L$. We did not
observe such anomalous behaviour for the Ising system.
Work is in progress on this issue.

(3) It is interesting to note that when the range $r$ of the long-range
    interaction varies from site to site and takes any integer value
    chosen randomly between $2$ and $L$, the persistence behaviour almost
    vanishes \cite{pratap}. Since we have found the persistence behaviour
    to remain upto time $\tau_1 \sim r^2$, for a given value of $r$ we conclude
    that the random nature of $r$ tends to remove the persistence behaviour.

\begin{acknowledgements}
The work of one author (AKC) was supported by UGC fellowship.
We also acknowledge the financial support from UPE (Computational
Group) grant for computational
facility.\\\end{acknowledgements}


\begin{references}

\bibitem{Majumdar}
S.N. Majumdar, Curr. Sci. {\bf 77}, 370 (1999). 

\bibitem{Ray}
P. Ray, Phase Transitions, {\bf 77}, 563 (2004). 

\bibitem{derrida1} B.~Derrida, V.~Hakim and V.~Pasquier, Phys. Rev. Lett. 
{\bf 75}, 751 (1995); J. Stat. Phys. {\bf 85}, 763 (1996). 

\bibitem{stauffer} D.~Stauffer, J Phys. A: Math. Gen. {\bf 27}, 5029 (1994).

\bibitem{derrida2} B.~Derrida, A.J.~Bray, and C. Godreche, J. Phys. A 
{\bf 27}, L357 (1994).

\bibitem{jain} S. Jain, Phys. Rev. E {\bf 59}, R2493 (1999).

\bibitem{bray} S.J.~O'Donoghue and A.J.~Bray, Phys. Rev. E {\bf 64}, 041105 
(2001) (see below Eq.(15) and Fig. 8). 

\bibitem{redner} S. Redner, in {\em Nonequilibrium Statistical Mechanics in 
One Dimension}, Ed. V. Privman, Cambridge University Press, 1997 (p. 3).

\bibitem{pratap} P.K. Das and P. Sen, Eur. Phys. J. B {\bf 47}, 391 (2005). 

\end{references}
\end{document}